\documentclass[twocolumn,prl,showpacs,amsmath,amssymb,superscriptaddress]{revtex4}
\usepackage{graphicx,bm,dcolumn,amsmath,amssymb,amsfonts}
\newcommand{\Hop}{\hat{H}}

\newcommand{\psiop}{\hat{\psi}}
\newcommand{\vecr}{\bm{r}}

\newcommand{\Tr}{{\rm Tr}}
\newcommand{\Gop}{\hat{G}}

\newcommand{\pdt}{\partial_{t}}
\newcommand{\pdx}{\partial_{x}}
\begin{document}

\title{Theory of heterotic SIS Josephson junctions between single- and multi-gap superconductors}

\affiliation{
CCSE, Japan Atomic Energy Agency, 
6-9-3 Higashi-Ueno Taito-ku, Tokyo 110-0015, Japan}
\affiliation{
Institute for Materials Research, Tohoku University, 2-1-1 Katahira
Aoba-ku, Sendai 980-8577, Japan} 
\affiliation{
CREST(JST), 4-1-8 Honcho, Kawaguchi, Saitama 332-0012, Japan}
\affiliation{
JST, TRIP, Sambancho Chiyoda-ku, Tokyo 102-0075, Japan}
\author{Yukihiro Ota}
\affiliation{
CCSE, Japan Atomic Energy Agency, 
6-9-3 Higashi-Ueno Taito-ku, Tokyo 110-0015, Japan}
\affiliation{
CREST(JST), 4-1-8 Honcho, Kawaguchi, Saitama 332-0012, Japan}
\author{Masahiko Machida}
\affiliation{
CCSE, Japan Atomic Energy Agency, 
6-9-3 Higashi-Ueno Taito-ku, Tokyo 110-0015, Japan}
\affiliation{
CREST(JST), 4-1-8 Honcho, Kawaguchi, Saitama 332-0012, Japan}
\affiliation{
JST, TRIP, Sambancho Chiyoda-ku, Tokyo 102-0075, Japan}
\author{Tomio Koyama}
\affiliation{
Institute for Materials Research, Tohoku University, 
2-1-1 Katahira Aoba-ku, Sendai 980-8577, Japan}
\affiliation{
CREST(JST), 4-1-8 Honcho, Kawaguchi, Saitama 332-0012, Japan}
\author{Hideki Matsumoto}
\affiliation{
Institute for Materials Research, Tohoku University, 
2-1-1 Katahira Aoba-ku, Sendai 980-8577, Japan}
\affiliation{
CREST(JST), 4-1-8 Honcho, Kawaguchi, Saitama 332-0012, Japan}
\date{\today}

\begin{abstract}
Using the functional integral method, we construct a theory of heterotic
 SIS Josephson junctions between single- and two-gap superconductors. 
The theory predicts the presence of in-phase and out-of-phase
 collective oscillation modes of superconducting phases. 
The former corresponds to the Josephson plasma mode whose frequency is
 drastically reduced for $\pm$ s-wave symmetry, and the latter
 is a counterpart of Leggett's mode in Josephson junctions.
We also reveal that the critical current and the Fraunhofer pattern
 strongly depend on the symmetry type of the two-gap superconductor. 
\end{abstract}

\maketitle

The Josephson effect is one of the most drastic phenomena
in superconductivity~\cite{Tinkham2004}. 
Cooper pairs can tunnel through an insulating barrier in a
non-dissipative manner.  
This particular feature has attracted tremendous attention of 
not only physicists but also device engineers.

Very recently, multi-gap superconductors have been revisited since 
the discovery of an iron-based high-$T_{{\rm c}}$
superconductor~\cite{Kamihara;Hosono:2008,Takahashi2008,Ren;Zhao:2008}. 
In contrast to cuprate high-$T_{{\rm c}}$ superconductors, 3-$d$ electrons
on the iron atom form multi-bands whose Cooper pairs condense into a
multi-gap superconducting state. 
The angle resolved photoemission spectroscopy has reported that 
each of multiple disconnected Fermi surfaces is fully
gapped~\cite{Ding2008} and other experiments have also supported the gapful
features~\cite{fullgap}. 
On the contrary, the nuclear magnetic resonance have shown typical 
gapless features~\cite{Nakai;Hosono;2008}.

\begin{figure}[bp]
 \scalebox{0.32}[0.32]{\includegraphics{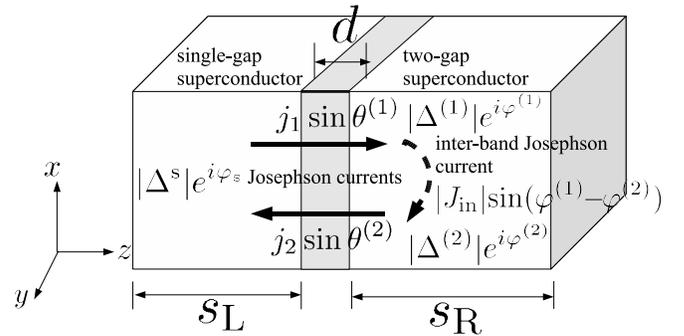}}
\caption{A schematic figure of the present heterotic junction system. The left electrode is
 a single-gap superconductor, and the right electrode a two-gap superconductor.}
\label{fig:junctions} 
\end{figure}

In order to compromise the controversy, the presence of $\pm$ s-wave
gaps on the disconnected Fermi surfaces has been  
proposed~\cite{Mazin;Du:2008,Kuroki;Aoki:2008,Nagai;Machida:2008}. 
The essence of the $\pm$ s-wave symmetry is a sign change between
different s-wave order parameters.  
This is expected to bring about novel behaviors 
in phase interference effects.
In particular, Josephson effects in SIS junction 
between the single- and the $\pm$ s-wave multi-gap superconductors 
as schematically illustrated in Fig.~\ref{fig:junctions} drastically
reflect the sign change. 
In this Letter, we focus on such a heterotic SIS junction
and clarify peculiar Josephson effects.
We have three main results, i.e., the drastic reduction of (i) the Josephson
plasma frequency, (ii) the critical current, and (iii) the Fraunhofer
pattern visibility.  
The $\pm$ s-wave symmetry leads to a cancellation between the two Josephson
currents which arise from the two tunneling channels in this system.

In the proposed junction as shown in Fig.~\ref{fig:junctions}, the left
(right) electrode is a single-(two-)gap superconductor
with the width $s_{{\rm L}}$ ($s_{{\rm R}}$). 
The insulator width and the dielectric constant are $d$ and
$\epsilon$, respectively. 
The current and the magnetic field are applied along $z$- and
$y$-direction, respectively. 
Similar situations were also examined from other
viewpoints~\cite{others}.

The system's Hamiltonian 
\(
\Hop 
= \int_{{\rm R}}d^{3}\vecr (
\mathcal{\Hop}^{(1)}_{{\rm R}}
+\mathcal{\Hop}^{(2)}_{{\rm R}}
+\mathcal{\Hop}^{{\rm pair}}_{{\rm R}}
)
+ \int_{{\rm L}}d^{3}\vecr (
\mathcal{\Hop}^{{\rm s}}_{{\rm L}}
+\mathcal{\Hop}^{{\rm pair}}_{{\rm L}}
) + \Hop_{{\rm T}}
\), 
where 
\(
\mathcal{\Hop}^{(i)}_{\rm R}
\) describes the kinetic energy of the $i$-th band electrons
$\psiop^{(i)}_{\sigma}$ in the right electrode.  
The pairing term in the right electrode 
\(
 \mathcal{\Hop}^{{\rm pair}}_{\rm R}
= -g_{1}
\psiop^{(1)\,\dagger}_{\uparrow} \psiop^{(1)\,\dagger}_{\downarrow}
\psiop^{(1)}_{\downarrow} \psiop^{(1)}_{\uparrow}
-g_{2}
\psiop^{(2)\,\dagger}_{\uparrow} \psiop^{(2)\,\dagger}_{\downarrow}
\psiop^{(2)}_{\downarrow} \psiop^{(2)}_{\uparrow}
-g_{12}
(
\psiop^{(1)\,\dagger}_{\uparrow} \psiop^{(1)\,\dagger}_{\downarrow}
\psiop^{(2)}_{\downarrow} \psiop^{(2)}_{\uparrow}
+ \text{h.c.})
\)~\cite{SGB2002}, 
where $g_{i}>0$, and the inter-band interaction can be either
attractive (i.e., $g_{12}>0$) or repulsive (i.e., $g_{12}<0$). 
The left electrode is described by
\(
\mathcal{\Hop}^{{\rm s}}_{{\rm L}} 
+ \mathcal{\Hop}^{{\rm pair}}_{{\rm L}}
\), where 
\(
\mathcal{\Hop}^{{\rm s}}_{{\rm L}}
\) is the kinetic energy and 
\(
\mathcal{\Hop}^{{\rm pair}}_{{\rm L}}
=-g_{{\rm s}}
\psiop^{{\rm s}\,\dagger}_{\uparrow} \psiop^{{\rm s}\,\dagger}_{\downarrow}
\psiop^{{\rm s}}_{\downarrow} \psiop^{{\rm s}}_{\uparrow}
\), in which $g_{{\rm s}}>0$. 
The tunneling Hamiltonian 
\(
\Hop_{{\rm T}} = \Hop_{{\rm T}}^{(1)} + \Hop_{{\rm T}}^{(2)}
\), 
where 
\(
\hat{H}_{{\rm T}}^{(i)}
\)
means the tunneling 
between the electrons in the left side and the $i$-th band electrons in
the right side. 
Using the imaginary time functional integral
method~\cite{Simanek1994,MKTT2000}, the effective action with respect to
the order parameters $\Delta^{(i)}$ and $\Delta^{{\rm s}}$ is given by 
\(
S_{\rm eff}
= 
\int^{\hbar\beta}_{0} \!\!\!
d\tau 
[\!
\int_{{\rm R}} d^{3}\vecr
(g_{2}|\Delta^{(1)}|^{2}/g
+ g_{1}|\Delta^{(2)}|^{2}/g
+V_{\rm in})
+\int_{{\rm L}} d^{3}\vecr
|\Delta^{{\rm s}}|^{2}/g_{{\rm s}}
]
- \Tr \ln \Gop_{0} - \Tr \ln\Gop^{-1}
\), 
where $\beta$ is the inverse temerature and 
\(
V_{{\rm in}} 
= -g_{12}(\Delta^{(1)\ast}\Delta^{(2)} + \text{c.c.})/g 
\). 
We assume here that $g\equiv g_{1}g_{2} - g_{12}^{2}>0$~\cite{SGB2002}.  
The Green functions for the non-interacting system $\Gop_{0}$ and the
total system $\Gop$ are $6\times 6$ matrices.  
We do not write their explicite expressions here, but those consists of
$4\times 4$ for the two-gap and $2\times 2$ for the
single-gap superconductors~\cite{Simanek1994,MKTT2000}.
The inter-band Josephson coupling term $V_{{\rm in}}$ is rewritten as 
\(
V_{{\rm in}}=-2(g_{12}/g)|\Delta^{(1)}|\,|\Delta^{(2)}| 
\cos (\varphi^{(1)}-\varphi^{(2)})
\), in which 
\(
\Delta^{(i)} = |\Delta^{(i)}|e^{i\varphi^{(i)}}
\). 

Based on the standard procedure~\cite{Simanek1994,MKTT2000}, the effective
Lagrangian density of the superconducting phases on $zx$-plane in the real time
formalisum is given by
\begin{eqnarray}
 \mathcal{L}_{{\rm eff}} 
&=&
\frac{s_{{\rm L}}}{8\pi \mu^{2}} a_{0}^{2}
+
\sum_{i=1}^{2} 
\frac{s_{{\rm R}}}{8\pi \mu^{(i)\,2}}
a_{0(i)}^{2} 
-\frac{s_{{\rm L}}}{8\pi\lambda^{2}}
a_{x}^{2} \nonumber \\
&&
-
\sum_{i=1}^{2}
\frac{s_{{\rm R}}}{8\pi\lambda^{(i)\,2}}
a_{x(i)}^{2}
-V_{{\rm J}}
+ \mathcal{L}_{{\rm EM}}, 
\label{eq:efflag} 
\end{eqnarray} 
where
\begin{eqnarray}
V_{{\rm J}}
&=&
-\frac{\hbar j_{1}}{e^{\ast}}\cos\theta^{(1)}
-\frac{\hbar j_{2}}{e^{\ast}}\cos\theta^{(2)} 
-\frac{\hbar J_{{\rm in}}}{e^{\ast}}\cos\varphi
\label{eq:jenergy}, \\
\mathcal{L}_{{\rm EM}}
&=&
\frac{\epsilon d}{8\pi} 
E_{{\rm RL}}^{z\,2} 
- 
\frac{d}{8\pi}B_{{\rm RL}}^{y\,2}
\label{eq:em}, \\
 \theta^{(i)} 
&=& 
\varphi^{(i)} - \varphi_{{\rm s}}  
- \frac{e^{\ast}d}{\hbar c} A^{z}_{{\rm RL}}
\label{eq:inv_pd}, \\
\varphi
&=&\varphi^{(1)}-\varphi^{(2)}=\theta^{(1)}-\theta^{(2)}
\label{eq:def_rltp},
\end{eqnarray} 
and note that  
\(
a_{0}
=
(\hbar/e^{\ast})\pdt\varphi_{{\rm s}}  + A_{{\rm L}}^{0}
\), 
\(
a_{x}
=
(\hbar c/e^{\ast}) \pdx\varphi_{{\rm s}} - A_{{\rm L}}^{x}
\), 
\(
a_{0(i)} 
= 
(\hbar/e^{\ast}) \pdt\varphi^{(i)} + A_{{\rm R}}^{0}
\),  
\(
a_{x(i)} 
=
(\hbar c/e^{\ast}) \pdx\varphi^{(i)} - A_{{\rm R}}^{x}
\), and 
\(
e^{\ast}=2e
\). 
The phase $\varphi_{{\rm s}}$ is defined as 
\(
\Delta^{{\rm s}} = |\Delta^{{\rm s}}|e^{i\varphi_{{\rm s}}}
\), and $j_{i}$ is the Josephson critical current between $i$-th and
single-band Cooper pairs. 
The charge screening length and the penetration depth on the left
(right) electrode are $\mu$ ($\mu^{(i)}$) and $\lambda$
($\lambda^{(i)}$), respectively. 
The last term in the gauge-invariant phase difference (\ref{eq:inv_pd})
is the $z$ component of the spatial averaged vector potential in the
insulator, defined as
\(
 A^{z}_{{\rm RL}} = d^{-1}\int_{-d/2}^{d/2} A^{z}(z)\, dz
\).
The electric and the magnetic fields in the insulator 
are defined as 
\(
E_{{\rm RL}}^{z} 
= -c^{-1} \pdt A^{z}_{{\rm RL}} -
d^{-1} (A^{0}_{{\rm R}} - A^{0}_{{\rm L}})
\) and 
\(
B_{{\rm RL}}^{y} = d^{-1}(A^{x}_{{\rm R}}-A^{x}_{{\rm L}})-\pdx
 A^{z}_{{\rm RL}}
\), respectively. 
Here, let us focus on the Josephson coupling energy (\ref{eq:jenergy}). 
The first and the second terms are the ordinary Josephson coupling terms,
while the third term corresponds to the inter-band Josephson coupling
energy and $|J_{{\rm in}}|$ is proportional to $|(g_{12}/g)s_{{\rm R}}|$. 
One finds that $J_{{\rm in}}$ is positive (negative) if $g_{12}>0$
($g_{12}<0$). 
If $s_{{\rm L}}$ and $s_{{\rm R}}$ are much larger than $d$, then it allows
us to regard $|J_{{\rm in}}| \gg j_{1},\,j_{2}$. 

From Eq.~(\ref{eq:efflag}), we have the Euler-Lagrange 
equations with respect to $A^{0}_{l}$ and $A^{x}_{l}$ as follows,   
\begin{eqnarray}
 \frac{\bar{\alpha}}{\alpha_{1}} \pdt\theta^{(1)} 
+ \frac{\bar{\alpha}}{\alpha_{2}} \pdt\theta^{(2)} 
&=& 
C \frac{e^{\ast}d}{\hbar} E^{z}_{{\rm RL}}, 
\label{eq:genJRvolt}\\
 \frac{\bar{\eta}}{\eta_{1}} \pdx\theta^{(1)} 
+ \frac{\bar{\eta}}{\eta_{2}} \pdx\theta^{(2)} 
&=& L \frac{e^{\ast}d}{\hbar c} B^{y}_{{\rm RL}}
\label{eq:genJRmag},
\end{eqnarray}
where the dimensionless parameters in each electrode are defined as 
\(
\alpha = \epsilon \mu^{2}/s_{{\rm L}}d
\), 
\(
\alpha_{i} = \epsilon \mu_{i}^{2}/s_{{\rm R}}d
\), 
\(
\bar{\alpha}^{-1} = \alpha_{1}^{-1} + \alpha_{2}^{-1}
\), 
\(
\eta = \lambda^{2}/s_{{\rm L}}d 
\),   
\(
\eta_{i} = \lambda^{(i)\,2}/s_{{\rm R}}d
\), and 
\(
\bar{\eta}^{-1} = \eta_{1}^{-1} + \eta_{2}^{-1}
\). 
The magnitude of the electric (magnetic) field coupling is characterized
by $\alpha$ and $\alpha_{i}$ ($\eta$ and $\eta_{i}$)~\cite{MS2004}.  
The constants $C$ and $L$ are defined as 
\(
C = 1+ \alpha + \bar{\alpha}
\) 
and \(
L = 1+ \eta + \bar{\eta}
\), respectively. 
Equations (\ref{eq:genJRvolt}) and (\ref{eq:genJRmag})
correspond to the generalized Josephson relations~\cite{MKT1999}. 
The Euler-Lagrange equation with respect to 
$A^{z}_{{\rm RL}}$ gives the Maxwell equation,
\begin{eqnarray}
 \frac{e^{\ast}d}{\hbar c}\pdx B^{y}_{{\rm RL}}
&=& 
\sum_{i=1}^{2}\frac{1}{\lambda_{{{\rm J}}i}^{2}} \sin\theta^{(i)}
+\frac{\epsilon}{c^{2}}\frac{e^{\ast} d}{\hbar}\pdt E^{z}_{{\rm RL}}
\label{eq:Maxwell}, 
\end{eqnarray} 
where 
\(
\lambda_{{{\rm J}}i}^{-2} 
= 4\pi e^{\ast}d j_{i}/\hbar c^{2}
\). 
The first term on the right hand side of Eq.~(\ref{eq:Maxwell}) is the
summation of the Josephson current terms [Fig.~\ref{fig:junctions}].
Using Eqs.~(\ref{eq:genJRvolt})-(\ref{eq:Maxwell}), we obtain 
\begin{equation}
\sum_{i=1}^{2}
 \frac{C\bar{\eta}}{\eta_{i}}\pdx^{2}\theta^{(i)} 
=
\sum_{i=1}^{2}
\frac{CL}{\lambda_{{\rm J}i}^{2}}\sin\theta^{(i)}
+
\sum_{i=1}^{2}
 \frac{L\bar{\alpha}}{\alpha_{i}}\pdt^{2}\theta^{(i)}. 
\label{eq:eq_mp}
\end{equation} 
Next, from the Euler-Lagrange equations about 
$\varphi_{{\rm s}}$ and $\varphi^{(i)}$, we have 
\begin{eqnarray}
&&
\frac{\epsilon}{c^{2}}
\sum_{i=1}^{2}
\bigg[
(-1)^{i+1}\frac{CL}{\alpha_{i}}\pdt^{2}\theta^{(i)} 
+
\frac{(1+\alpha)\xi L }{\alpha_{i}}\pdt^{2}\theta^{(i)} 
\bigg] \nonumber \\
&=&
\sum_{i=1}^{2}
\bigg[
(-1)^{i+1}\frac{CL}{\eta_{i}}\pdx^{2}\theta^{(i)} 
+
 \frac{(1+\eta)\zeta C }{\eta_{i}}\pdx^{2}\theta^{(i)} 
\bigg] \nonumber \\
&&
- 
2CL \frac{4\pi e^{\ast}d}{\hbar c^{2}}
\frac{\partial V_{{\rm J}}}{\partial \varphi}
\label{eq:eq_rlt}, 
\end{eqnarray}
where the parameter $\xi$ ($\zeta$) means the difference of the 
magnitude of the electric (magnetic) field coupling between
the different superconducting bands as 
\(
\xi = (\alpha_{1}-\alpha_{2})/(\alpha_{1}+\alpha_{2})
\) and 
\(
\zeta = (\eta_{1}-\eta_{2})/(\eta_{1}+\eta_{2})
\). 

The present description is valid from very thin electrode junctions
($s_{{\rm L}} \sim \mu$ and $s_{{\rm R}}\sim \mu^{(i)}$) to conventional thick ones. 
In the latter case ($s_{{\rm L}}\gg\mu $ and $s_{{\rm R}} \gg \mu^{(i)}$), we can take an
approximate treatment, $\alpha \to 0$ and $\alpha_{i}\to 0$. 
Remark that the present paper highlight, i.e., particular features due
to $\pm$ s-wave is unchanged in this limit. 

\begin{figure}[tbp]
(a)\!\!\scalebox{0.652}[0.652]{\includegraphics{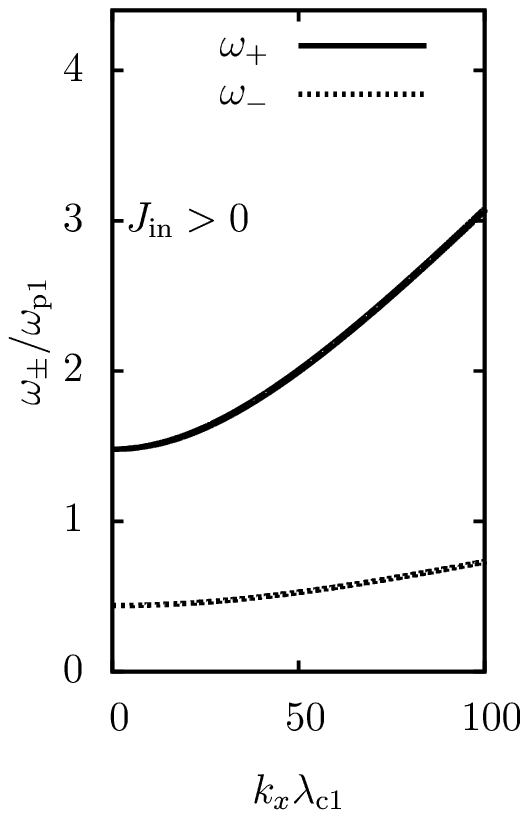}} 
(b)\!\!\scalebox{0.652}[0.652]{\includegraphics{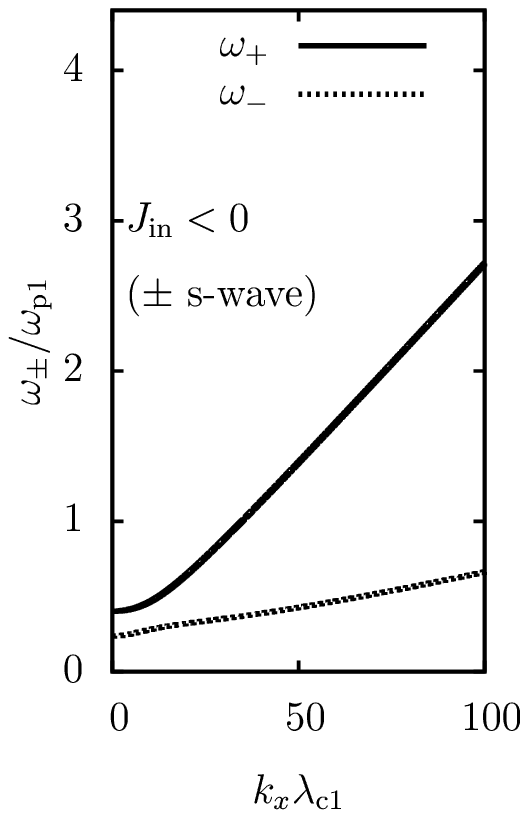}}
\caption{The dispersion relations for $\omega_{\pm}$. The
 solid (dotted) line is for $\omega_{+}$ ($\omega_{-}$). The
 parameters are set as follows: $\alpha=\alpha_{1}=10^{-3}$,
 $\alpha_{2}/\alpha_{1}=1.4$, $\eta=\eta_{1}=10^{3}$,
 $\eta_{2}/\eta_{1}=1.4$, $j_{2}/j_{1}=0.9$ and 
$|J_{{\rm in}}|/j_{1}=40.0$. The reduction of $\omega_{+}$ is
 observed for $J_{{\rm in}}<0$. (a) $J_{{\rm in}}>0$. (b) $J_{{\rm in}}<0$.}
\label{fig:dispersion} 
\end{figure}

Now, let us examine the collective modes involved in
Eqs.(\ref{eq:eq_mp}) and (\ref{eq:eq_rlt}).  
For this purpose, we linearize them around a stable point of $V$. 
First, we focus on $J_{{\rm in}}>0$. 
Then, a stable point for $(\theta^{(1)},\theta^{(2)})$ is $(0,0)$ and
the dispersion relations $\omega_{\pm}^{2}(k_{x})$ is given by 
\begin{equation}
 \omega_{{\pm}}^{2}(J_{{\rm in}}>0)
=
\frac{
X(\omega_{{\rm P}}^{2}+\omega_{{\rm L}}^{2}) 
-2\bar{\alpha}\xi C^{-1}D
\pm \sqrt{R}}{2(1-\xi^{2})},
\label{eq:plus_disp}
\end{equation} 
where
\begin{equation*}
\left\{
\begin{array}{l}
 \omega_{{\rm P}}^{2}
= C(\omega_{{\rm p}1}^{2} + \omega_{{\rm p}2}^{2})
[1 + L^{-1}(k_{x} /K)^{2}], \\
 \omega_{{\rm L}}^{2}
=
\bar{\alpha}X^{-1}
\{
4\nu_{{\rm in}}^{2}
+(\omega_{{\rm p}1}^{2} + \omega_{{\rm p}2}^{2})
[
1 + \bar{\eta}^{-1}Y (k_{x} /K)^{2}
]\}.
\end{array}
\right.
\end{equation*}
The Josephson plasma frequency associated with the Josephson current for
$\theta^{(i)}$, 
\(
 \omega_{{\rm p}i} 
= c/\sqrt{\epsilon}\lambda_{{{\rm J}}i}
\), while the pseudo Josephson-plasma frequency associated with the
inter-band Josephson current, 
\(
\nu_{{\rm in}} 
= \sqrt{
4\pi e^{\ast}d|J_{\rm in}|/\epsilon \hbar}
\). 
Note that 
\(
K^{2}
=c^{-2}\epsilon
(\omega_{{\rm p}1}^{2}+\omega_{{\rm p}2}^{2})
\), the dimensionless parameters $X$ and $Y$ are defined as 
\(
X = 1 - \xi^{2} C^{-1} (1+\alpha)
\) and   
\(
Y = 1 - \zeta^{2} L^{-1} (1+\eta)
\), respectively, and the quantities $D$ and $R$ are, respectively,   
\(
 D
= C(\omega_{{\rm p}1}^{2}+\omega_{{\rm p}2}^{2})
[-\delta + \zeta  L^{-1}(k_{x} /K)^{2}]
\) and 
\(
 R = X
\{
(1-\xi^{2})(\omega_{{\rm P}}^{2}-\omega_{{\rm L}}^{2})^{2}
+4\bar{\alpha}C^{-1} [D - \xi(\omega_{{\rm P}}^{2}+\omega_{{\rm L}}^{2})/2]^{2}
\}
\), where 
\(
\delta = (j_{1}-j_{2})/(j_{1}+j_{2})
\).  
When $\xi=\zeta=\delta=0$ (i.e., the superconducting
characters are perfectly equivalent between the two bands), we find that 
$\omega_{+}=\omega_{{\rm P}}$ and $\omega_{-}=\omega_{{\rm L}}$. 
We then notice that no term related to the inter-band Josephson coupling is
involved in the expression of $\omega_{{\rm P}}$.  
It indicates that the origin of $\omega_{{\rm P}}$ is irrelevant to the
motion of the relative phase $\varphi$. 
Then $\omega_{+}$ corresponds to the in-phase motion for $\theta^{(1)}$ and
$\theta^{(2)}$. 
On the other hand, the origin of $\omega_{-}$ is the
out-of-phase motion for $\theta^{(1)}$ and $\theta^{(2)}$.  
Next, we study another case of $J_{{\rm in}}<0$, in which 
$(\pi,0)$ is a stable point since $|J_{{\rm in}}| \gg j_{1},j_{2}$. 
Expanding $V$ around $(\pi,0)$, we have
\begin{equation}
 \omega_{\pm}^{2}(J_{\rm in}<0)
=
\frac{X(\omega_{{\rm P}}^{\prime\,2}+\omega_{{\rm L}}^{\prime\,2}) 
-2\bar{\alpha}\xi C^{-1}D^{\prime} \pm \sqrt{R^{\prime}}}
{2(1-\xi^{2})},
\label{eq:minus_disp}
\end{equation}
where 
\begin{equation*}
\left\{
\begin{array}{l}
 \omega^{\prime\,2}_{{\rm P}}
=
C(\omega_{{\rm p}1}^{2} + \omega_{{\rm p}2}^{2})
(|\delta| + L^{-1}(k_{x} /K)^{2} ), \\
 \omega^{\prime\,2}_{{\rm L}}
= 
\bar{\alpha}X^{-1}
[4\nu_{{\rm in}}^{2}
+(\omega_{{\rm p}1}^{2} + \omega_{{\rm p}2}^{2})
(
|\delta| + \bar{\eta}^{-1} Y (k_{x}/K)^{2}].
\end{array}
\right.
\end{equation*}
The quantities $D^{\prime}$ and $R^{\prime}$ are
\(
D^{\prime}
=C(\omega_{{\rm p}1}^{2}+\omega_{{\rm p}2}^{2})
(1+ \zeta L^{-1}(k_{x}/K)^{2} )
\) and 
\(
R^{\prime}
= X\{
(1-\xi^{2})(\omega_{{\rm P}}^{\prime\,2}-\omega_{{\rm L}}^{\prime\,2})^{2}
+4\bar{\alpha}C^{-1} 
[D^{\prime} - \xi(\omega_{{\rm P}}^{\prime\,2} - \omega_{{\rm L}}^{\prime\,2})/2]^{2}\}
\), respectively. 
Figures \ref{fig:dispersion}(a) and (b) show the typical dispersion
relations for $J_{{\rm in}}>0$ and $J_{{\rm in}}<0$, respectively.  
For both cases, the frequency of the out-of-phase mode $\omega_{-}$ is found to be
lower than the in-phase mode $\omega_{+}$ for an arbitrary value of $k$. 
Here, we take the limit $k_{x}\to 0$ in Eqs.~(\ref{eq:plus_disp}) and
(\ref{eq:minus_disp}) to explicitly evaluate the gap frequency for these
modes. 
As for $J_{{\rm in}}>0$, the leading order terms are,
respectively, given as 
\(
\omega_{+} \simeq (\omega_{{\rm p}1}^{2}+\omega_{{\rm p}2}^{2})^{1/2}
\) and 
\(
\omega_{-} \simeq 
(\alpha_{1}+\alpha_{2})^{1/2}\nu_{{\rm in}}
\) by regarding $\bar{\alpha}$ and $\alpha$ to be small. 
However, remark that we keep the term 
\(
\bar{\alpha}\nu_{{\rm in}}^{2}(\omega_{{\rm p}1}^{2} + \omega_{{\rm p}2}^{2})^{-1}
\) in the above evaluation, because $|J_{{\rm in}}| > j_{1},\,j_{2}$
even though $\bar{\alpha}$ is small.  
Similarly, when $J_{{\rm in}}<0$, we have 
\(
\omega_{+} \simeq |\omega_{{\rm p}1}^{2}-\omega_{{\rm p}2}^{2}|^{1/2}
\) and 
\(
\omega_{-} \simeq 
(\alpha_{1}+\alpha_{2})^{1/2}\nu_{{\rm in}}
\). 
The gap of $\omega_{+}$ is characterized by a superposition 
($J_{{\rm in}} > 0 $) or subtraction ($J_{{\rm in}} <0 $) between
$\omega_{{\rm p}1}$ and $\omega_{{\rm p}2}$.  
Thus, we refer $\omega_{+}$ to the Josephson plasma mode. 
We emphasize that the a signature of $\pm$ s-wave is the reduction of
plasma frequency [Fig.~\ref{fig:dispersion}].  
On the other hand, since the gap of $\omega_{-}$ is characterized by
$\nu_{{\rm in}}$ and $\alpha_{i}$, we find that it corresponds to the
gap of the Leggett's mode, which was derived as a collective mode
generated by the density fluctuation between two
superfluidities~\cite{Leggett1966}.  
Thus, it should be called Josephson-Leggett mode. 
The inter-band Josephson coupling and the charge density fluctuation
create the the mode. 
Conventionally, the first and the second terms in Eq.~(\ref{eq:efflag})
are fixed to be zero ($\alpha_{i}\to 0$) because the charge screening length is 
much smaller than the electrode size. 
Then, the effective Lagrangian density gives the standard Josephson relation 
\(
\pdt\theta^{(i)} = (e^{\ast}d/\hbar)E^{z}_{{\rm RL}}
\), 
resulting in
\( 
\pdt (\theta^{(1)}-\theta^{(2)}) =0
\). 
It means that the Josephson-Leggett mode becomes a gapless mode. 
In contrast, the mode $\omega_{{\rm P}}$ can still remain massive in
$\alpha_{i}\to 0$.  
The retainment of non-zero $\alpha_{i}$ is responsible for the finite
gap frequency of the Josephson-Leggett mode. 
The bulk Leggett's mode is normally embedded in the quasi-particle excitation
continuum~\cite{Blumberg;Karpinski:2007}, while the Josephson-Leggett mode
is more clearly and easily observable because the mode lies far beneath the gap
energy. 

\begin{figure*}[tbp]
(a)\!\!\scalebox{0.4}[0.4]{\includegraphics{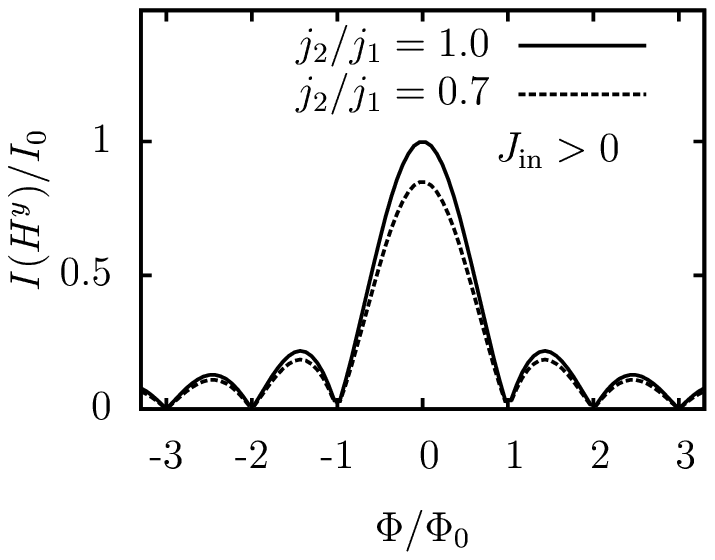}}\quad
(b)\!\!\scalebox{0.4}[0.4]{\includegraphics{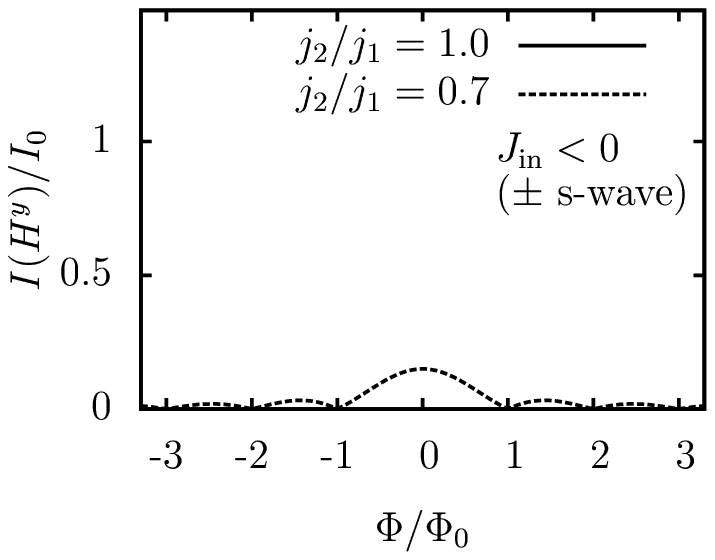}}\quad
(c)\scalebox{0.15}[0.15]{\includegraphics{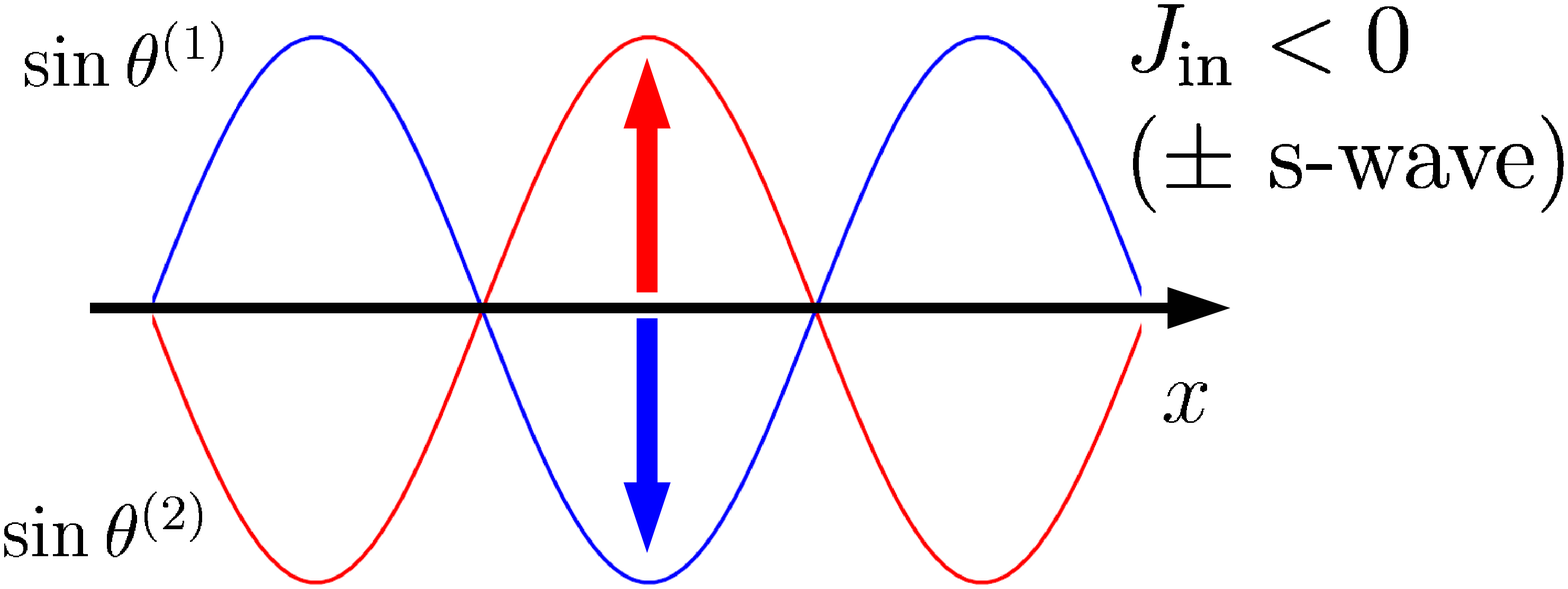}}
\caption{The current vs. the magnetic flux. The value of $I_{0}$ is
 $I(H^{y}=0)$ for $\varphi=0$. (a) $J_{{\rm in}}>0$ and $\varphi=0$. (b) 
$J_{{\rm in}}<0$ and $\varphi=\pi$. No current is observed when
 $j_{1}=j_{2}$. (c) The demonstration of the cancellation between two
 Josephson currents.} 
\label{fig:fd}
\end{figure*}
The remaining part of this Letter is devoted to basic Josephson effects. 
First, let us discuss the Josephson critical current $j_{{\rm c}}$. 
The bias current is assumed to be uniformly applied without the external
magnetic field.  
Namely, $\theta^{(1)}$ and $\theta^{(2)}$ are assumed to be unifrom along
the $x$-axis in Fig.~\ref{fig:junctions}. 
The bias current $I$ is added to the right hand side of Eq.~(\ref{eq:eq_mp}) with the
elimination of the spatial dependent terms, and $j_{{\rm c}}$ can be derived by
estimating the maximum threshold of $I$ under keeping a stationary
solution.  
The condition is given by
\begin{eqnarray}
I
&=&
j_{1}\sin\theta^{(1)}
+
j_{2}\sin\theta^{(2)},
\label{eq:balance_current} \\
0
&=&
-j_{1}\sin\theta^{(1)}
+ j_{2}\sin\theta^{(2)}
-J_{{\rm in}}\sin\varphi. 
\label{eq:in_current}
\end{eqnarray}
Equation (\ref{eq:balance_current}) means that $I$ coincides with the
sum of two Josephson currents between the electrodes, 
while Eq.~(\ref{eq:in_current}) is an internal current conservation law,
which gives a significant constraint on the critical current.  
When $J_{{\rm in}}>0$, the preferable choice of $\varphi$ is $0$. 
Equation (\ref{eq:in_current}) implies $(j_{1}-j_{2})\sin\theta^{(1)}=0$, because
$\theta^{(1)}=\theta^{(2)}$. 
This is always satisfied if $j_{1}=j_{2}$.  
Then, since $\theta^{(1)}$ can vary from $0$ to 
$2\pi$, $j_{{\rm c}}=j_{1}+j_{2}$. 
If $j_{1}\neq j_{2}$, then $\varphi$ can deviate from $0$ 
and $j_{{\rm c}}\le j_{1}+j_{2}$. 
On the other hand, when $J_{{\rm in}}<0$, $\varphi$ should be $\pi$. 
Equation (\ref{eq:in_current}) implies 
\(
(j_{1}+j_{2})\sin\theta^{(1)}=0
\). 
The only possible solution is $\theta^{(1)}=0$ and
$\theta^{(2)}=-\pi$, because $j_{1}+j_{2}\neq 0$. 
Thus, we find that the value of $j_{{\rm c}}$ is drastically reduced 
compared to the case of $J_{{\rm in}}>0$, e.g., $j_{{\rm c}}=0$ 
for the case of perfectly identical $\pm$ s-wave two-gap superconductivity.  

Next, we consider the Josephson effects in the presence of the external
magnetic field $H^y$. 
We focus on stationary solutions, i.e., we drop the temporal
terms of $\theta^{(i)}$.  
According to Eq.~(\ref{eq:genJRmag}), 
\(
 \theta^{(1)}(x)
= kx + \theta_{0} + (\bar{\eta}/\eta_{2}) \varphi(x) 
\) and 
\(
 \theta^{(2)}(x)
= kx + \theta_{0} - (\bar{\eta}/\eta_{1})\varphi(x)
\), 
where \(k=L(e^{\ast}d/\hbar c)H^{y}\) and $\theta_{0}\in[0,2\pi)$ is an
integral constant.   
The observed current is then given by
\(
 I(H^{y},\theta_{0}) 
= \int_{-L_{x}/2}^{L_{x}/2} 
\left[
j_{1}\sin \theta^{(1)}(x) + j_{2}\sin\theta^{(2)}(x)
\right] dx
\).  
Hereafter, we assume that $\varphi(x)$ is spatially uniform. 
Taking account of $|J_{{\rm in}}| > j_{1},\,j_{2}$, we should have 
\(
0 \approx -J_{{\rm in}}\sin\varphi
\) from Eq.~(\ref{eq:eq_rlt}). 
When $J_{{\rm in}}>0$ (i.e., $\varphi=0$), the magnetic field dependece
of the current is given by 
\(
 I(H^{y},\theta_{0})
=
L_{x}\sin\theta_{0} [
(j_{1}+j_{2})(\Phi_{0}/\pi\Phi)\sin(\pi\Phi/\Phi_{0})]
\), where $\Phi_{0}=2\pi\hbar c/e^{\ast}$ and $\Phi = LH^{y}dL_{x}$. 
Then, 
\(
I(H^{y})\equiv \max_{\theta_{0}}|I(H^{y},\theta_{0})|
= L_{x}(j_{1}+j_{2}) |(\Phi_{0}/\pi\Phi)\sin(\pi\Phi/\Phi_{0})|
\). 
As a result, we obtain the ordinary Fraunhofer diffraction pattern as a
function of the magnetic flux $\Phi$ [Fig.~\ref{fig:fd}(a)].  
The maximun value of $I(H^{y})$ is the sum of two
Josephson currents as
\(
I(0) = L_{x}(j_{1}+j_{2})
\). 
We also observe that the net current conventionally vanishes when 
$\Phi = \kappa\Phi_{0}$ ($\kappa\in \mathbb{N}$). 
In contrast, as for $J_{{\rm in}}<0$ (i.e., $\pm$ s-wave), 
the current is given by
\(
 I(H^{y})
= L_{x} 
|j_{1}-j_{2}| |(\Phi_{0}/\pi\Phi) \sin(\pi\Phi/\Phi_{0})|
\). 
If $j_{1} = j_{2}$, the Fraunhofer diffraction pattern completely
disappears. 
When $j_{1}\neq j_{2}$, the pattern is observable except for
$\Phi=\kappa\Phi_{0}$, but the maximum value becomes unexpectedly small
[Fig.~\ref{fig:fd}(b)].   
The situation at $j_{1}=j_{2}$ is schematically displayed in
Fig.~\ref{fig:fd}(c).  
The Josephson currents for $\theta^{(1)}$ and $\theta^{(2)}$ cancel out
each other. 

Finally, let us discuss how to experimentally confirm the theoretical
predictions. 
We point out that the maximum Josephson current can be estimated 
from the normal state resistance based on Ambegaokar-Baratoff
relation~\cite{Simanek1994} under an assumption $J_{{\rm in}}>0$. 
If the measured $j_{{\rm c}}$ is significantly reduced from the one estimated
above, then $J_{{\rm in}}<0$, i.e., $\pm$ s-wave symmetry is concluded. 

In summary, we microscopically derived an effective Lagrangian density
of the SIS Josephson junction between single- and two-gap superconductors and
examined the collective modes, the critical current, and the Fraunhofer
pattern.  
We found that these properties are considerably affected by the type of
the pairing symmetry of the two-gap superconductor. 
We conclude that the heterotic junction is useful to identify directly a
symmetry of two-gap superconductors. 

The authors (Y.O. and M.M) wish to acknowledge valuable discussion with
H. Aoki, S. Shamoto, Y. Ohashi, D. Inotani, N. Hayashi, Y. Nagai, S. Yamada, 
H. Nakamura, M. Okumura, and N. Nakai. 
M.M. specially thanks H. Fukuyama for his illuminating comments. 
The work was partially supported by Grant-in-Aid for Scientific Research
on Priority Area ``Physics of new quantum phases in superclean
materials'' (Grant No. 20029019) from the Ministry of Education,
Culture, Sports, Science and Technology of Japan. 
M.M. is supported by JSPS Core-to-Core
Program-Strategic Research Networks, ``Nanoscience and Engineering in
Superconductivity''.


\begin{thebibliography}{99}
\bibitem{Tinkham2004}
M. Tinkham, 
{\it Introduction to Superconductivity} 
(Dover, New York, 2004) 2nd ed.
\bibitem{Kamihara;Hosono:2008}
Y. Kamihara, {\it et al.}, J. Am. Chem. Soc. {\bf 130}, 3296 (2008).
\bibitem{Takahashi2008}
H. Takahashi, {\it et al.}, Nature {\bf 453}, 376 (2008).
\bibitem{Ren;Zhao:2008}
Z.-A. Ren, {\it et al.}, Chin. Phys. Lett. {\bf 25}, 2215 (2008).
\bibitem{Ding2008}
H. Ding, {\it et al.}, Euro. Phys. Lett. {\bf 83}, 47001 (2008).
\bibitem{fullgap}
A. Kawabata, {\it et al.}, 
J. Phys. Soc. Jpn. {\bf 77}, 103704 (2008); 
K. Hashimoto, {\it et al.}, 
Phys. Rev. Lett. {\bf 102}, 017002 (2009).
\bibitem{Nakai;Hosono;2008}
Y. Nakai, {\it et al.}, J. Phys. Soc. Jpn. {\bf 77}, 073701 (2008).
\bibitem{Mazin;Du:2008}
I. I. Mazin, {\it et al.}, 
Phys. Rev. Lett. {\bf 101}, 057003 (2008).
\bibitem{Kuroki;Aoki:2008}
K. Kuroki, {\it et al.}, 
Phys. Rev. Lett. {\bf 101}, 087004 (2008).
\bibitem{Nagai;Machida:2008}
Y. Nagai, {\it et al.}, 
New. J. Phys. {\bf 10}, 103026 (2008).
\bibitem{others}
D. F. Agterberg, E. Demler, and B. Janko, 
Phys. Rev. B {\bf 66}, 214507 (2002); 
T. K. Ng and N. Nagaosa, arXiv:0809.3343 (2008); 
D. Inotani and Y. Ohashi, arXiv:0901.1718 (2009);
J. Linder, I. B. Sperstad, and A. Sudb\o, arXiv:0901.1895 (2009).
\bibitem{SGB2002}
S. G. Sharapov, V. P. Gusynin, and H. Beck, 
Euro. Phys. J. B {\bf 30}, 45 (2002).
\bibitem{Simanek1994}
E. $\check{{\rm S}}$im$\acute{{\rm a}}$nek, 
{\it Inhomogeneous Superconductors: Granular and Quantum Effects} 
(Oxford University Press, New York, 1994).
\bibitem{MKTT2000}
M. Machida, {\it et al.}, 
Physica C {\bf 331} 85 (2000).
\bibitem{MS2004}
M. Machida and S. Sakai, 
Phys. Rev. B {\bf 70}, 144520 (2004). 
\bibitem{MKT1999}
M. Machida, T. Koyama, and M. Tachiki, 
Phys. Rev. Lett. {\bf 83}, 4618 (1999).
\bibitem{Leggett1966}
A. J. Leggett, 
Prog. Theor. Phys. {\bf 36}, 901 (1966).
\bibitem{Blumberg;Karpinski:2007}
G. Blumberg, {\it et al.}, 
Phys. Rev. Lett. {\bf 99}, 227002 (2007).
\end{thebibliography}
\end{document}